\documentclass[twoside]{article}
\usepackage{curves,fleqn,espcrc2}

\usepackage{graphicx}

\newcommand{\AmS}{{\protect\the\textfont2
  A\kern-.1667em\lower.5ex\hbox{M}\kern-.125emS}}

\title{Some universal properties of the string breaking}

\author{Ferdinando Gliozzi\address{Dipartimento di Fisica Teorica
        dell'Universit\`a di Torino, and \\
       Istituto Nazionale di Fisica Nucleare, via P.Giuria 1, 
       I-10125 Torino,Italy} }      
\begin{document}
\def\de{\partial}
\def\oh{\frac{1}{2}}
\newcommand\hb{\hbar}
\newcommand{\bra}{\langle}
\newcommand{\ket}{\rangle}
\newcommand{\um}{\frac 12}
\newcommand{\eq}{\begin{equation}}
\newcommand{\en}{\end{equation}}
\begin{abstract}

In  gauge systems coupled to  matter, the static potential 
flattens out at a  scale where the confining string breaks  by formation
 of a dynamical pair of particles. Surprisingly, such a  breaking
is invisible in  Wilson loops even when the inter-charge separation is
 much larger than the flattening scale. Observing string breaking 
also requires using different operators. A known string mechanism
provides us with a simple explanation, leading  to the area law for 
large Wilson loops, as observed in most gauge models coupled to 
whatever kind of matter.
It is also pointed out that in a simple 3D $Z_2$ gauge-Higgs model, once 
reformulated in terms of Fortuin Kasteleyn clusters, this peculiar 
behaviour of the Wilson operator can be ascribed to non-trivial
linking  of a percolating cluster to the Wilson loop. Some numerical
tests on this model are also presented.   

\vspace{1pc}
\end{abstract}

\maketitle

\section{INTRODUCTION}
In  gauge theories coupled to  matter (quarks or Higgs fields in the
fundamental representation)
the string breaking is the phenomenon  of static potential flattening
at large distances due to the screening of the sources
produced by  pair creation.
A similar screening is expected also in pure
Yang-Mills theory for sources in the adjoint representation, 
where the role of charged matter is played by the gluons ($ adjoint~ string$).

The coupled systems of this kind studied up to now reveal a
surprising, general phenomenon: the string breaking is invisible in the
Wilson loops, namely, the area law continues to prevail 
 in full QCD \cite{qcd1},\cite{qcd2} and in 
pure Yang Mills theory with adjoint sources \cite{adj} even at
distances where the static charges are completely screened.
In other terms there is no possibility to extract the true large
distance behaviour of the static potential using {\sl only} the Wilson loops. 
On the contrary, neat
descriptions of potential flattening have been obtained in studies
where the basis of the operators has been enlarged in order to gain
a better ground state overlap.  In this way it has been observed the
breaking of the confining string between colour sources in fundamental
representation in Higgs models \cite{ks}, in QCD \cite{milc} and the 
breaking of the adjoint string \cite{adjb}. 

 A possible explanation for this unexpected
 finding is that the available range of the Wilson loops of present
 studies is not wide enough and that for larger sizes 
 the overlap of the Wilson operators to the broken string state 
 could become visible. This however would imply that the
 presently observed area law obeyed by the Wilson loops, which has
 been tested for a wide range even in QCD \cite{qcd2},
 should  be broken at a larger scale of doubtful physical significance.    

There is another explanation, based on the effective string picture of
confinement and supported by a well known string mechanism \cite{ad}, 
which points in the opposite direction\cite{gp}. According to such a
picture, the area law describes the {\sl asymptotic, universal}
behaviour of the Wilson loops in most (perhaps all) gauge theories in
the  confined phase, coupled to whatever kind of matter. Any observed
non-vanishing overlap of the Wilson operator to the broken
string state should be a decreasing function of the loop size, going 
eventually to zero in the IR limit. If this prediction will pass the 
numerical tests of the future lattice simulations, one can envisage
using the asymptotic area law of the Wilson loops for an unambiguous
definition of the confining phase and string tension even when the
static potential flattens because of sea quarks or other 
dynamical matter. This will be discussed in section 2.

There is one case, at least, where the area law of the Wilson loops can
be proved without appealing to any effective  theory: In the 
 $3D$ $Z_2$ gauge Higgs model it can be shown that the very  
confining mechanism which is operating in pure $Z_2$ gauge theory
survives to the addition of $Z_2$ charged matter. In particular, for a
simple topological reason the Wilson loop continues to obey the area
law. This will be discussed in section 3, where a very efficient
numerical method is used to test the behaviour of large Wilson
loops.

\section{EFFECTIVE STRING BREAKING } 
The lack of visible effects of string breaking in the Wilson loop and
its manifestation in other operators have a simple
explanation in terms of the effective string picture of the confining
phase of any gauge theory.
 According to such a  description, the expectation value of a
 rectangular Wilson loop can be represented as 
\eq 
\langle W(R,T)\rangle\propto  e^{-F_o}~,
\label{one}
\en
where $F_o$ is the free energy of a 2D model describing the
normal modes of vibration of the string world-sheet bounded by the 
rectangle $R\times T$. This assumption yields, in the IR limit,
\eq
e^{-F_o}\sim e^{-\sigma R T-p(R+T)}
\left[\frac{\sqrt{R}}{q^{\frac1{24}}\prod_{n=1}^\infty(1-q^n)}
\right]^{\frac{d-2}{2}}
\label{asymp}
\en
where $q\equiv \exp(-2\pi T/R)$, $\sigma$ is the string tension and
$d$ the space-time dimension.
Adding dynamical matter fields  induces the formation of holes of 
any size in the world-sheet, hence eq.(\ref{one}) is replaced by a 
loop expansion:
\eq
\langle W\rangle\propto e^{-F_o} +\sum_{\rm one~ hole} e^{-F_1}+
\sum_{\rm two~ holes}e^{-F_2}+\dots
\label{more}
\en

\begin{center}
\begin{picture}(130,50)(5,-10)
\multiput(20,15)(65,0){3}{$+$}
\closecurve(40,15,50,25,60,10)\closecurve(130,12,125,30,140,10)
\closecurve(107,10,108,25,122,10)
\thicklines
\multiput(-32,0)(65,0){3}{\framebox(45,33)}
\multiput(-2,43)(65,0){3}{$\updownarrow$}
\end{picture}
\end{center}
\noindent
where the loops inside the Wilson rectangle represent pair created particles. 
What we can say about the sum of this series? A simple solvable matrix 
model, describing a free string with holes of any size, 
 suggests that there are two phases: 

In the {\sl normal} phase  this sum
is dominated by configurations with small holes. Their size
does not depend on that of the Wilson loop: Larger Wilson loops
have more holes. These holes do not influence the area law and their 
effect can be absorbed into a renormalization of the string tension,
according to a well known string mechanism \cite{ad}.

In the other ({\sl tearing}) phase the mean size of the holes 
increases with that of the Wilson loop. In this case  asymptotic 
screening should be visible also in the Wilson loops.

The observed poor overlap of the Wilson operator with
the broken string state suggests that in all the known cases 
the confining string belongs to the normal phase.
In this phase large Wilson loops behave exactly like
in the pure Yang Mills theory. There are however  finite size effects
which are worth mentioning. As a matter of fact, different 
phases are completely separated only in the thermodynamic limit. 
The expectation value of {\sl finite} Wilson loops  in the normal
phase receives a non-vanishing  contribution of configurations typical 
of the tearing phase, of course. 
Thus the overlap of a \underline{finite} Wilson loop to the
broken string state cannot be exactly zero, but should be a decreasing
function of its size, going to zero in the infinite size limit.
In order to gain an intuitive understanding of what is going on, 
imagine to cut a rectangular Wilson loop $R\times T$ at a given ``time''
(the dashed line in the following drawing)
\vskip 1. cm        
\begin{center}
\begin{picture}(130,50)(5,-10)

\multiput(-18,45)(20,0){9}{\oval(10,5)}
\multiput(-23,35)(22,0){8}{\oval(8,5)}
\multiput(-15,25)(24,0){7}{\oval(8,8)}
\multiput(-16,14)(18,0){10}{\oval(8,5)}
\multiput(-24,2)(42,0){4}{\oval(9,7)}
\multiput(-3,2)(43,1){4}{\oval(5,8)}
\multiput(140,30)(21,0){1}{\oval(5,5)}
\multiput(140,3)(21,0){1}{\oval(5,5)}
\multiput(20,-13)(0,8){10}{\line(0,1){5}}
\put(90,-17){\vector(1,0){60}}
\put(55,-17){\vector(-1,0){80}}
\put(160,18){R}
\put(163,30){\vector(0,1){18}}
\put(163,10){\vector(0,-1){16}}
\put(70,-18){T}
\thicklines
\multiput(-30,-5)(70,0){1}{\framebox(182,55)}
\end{picture}
\end{center}
The above naive picture of the string world-sheet riddled with holes 
suggests that the Wilson operator has an important overlap with a 
multi-meson state $\vert n\rangle$ where the mean number $n$ of
particles is a growing function of the distance $R$ between the static sources.  

The above reasoning does not imply that there is no flattening in the static 
potential.  The observed screening of  sources is a 
mixing phenomenon. It is simply due to the possibility, in presence 
of charged matter, to construct other operators with a sufficient
overlap to the ground state even for large separations between the static
source \cite{ks}, \cite{adjb}.  Applying the loop expansion (\ref{more}) to 
operators made of disconnected pieces leads to a contribution with a
large overlap with the broken string state. This very
argument also explains why \cite{gp} in the QCD at finite temperature 
the string breaking is neatly visible already in the Polyakov correlator
\cite{ft}.

\section{3D $Z_2$ GAUGE-HIGGS MODEL}
The action of a  3D $Z_2$ gauge theory coupled to a 
charged matter field can be written as 
\eq
S(\beta_G,\beta_I)=\beta_I\sum_{\langle ij\rangle}
\sigma_i U_{ij}\sigma_j+
\beta_G\sum_{plaq.}U_{\bar\sqcup}
\en
where both the link variable $U_{ij}\equiv U_\ell$  and the matter
field $\sigma_i$  take values $\pm1$ and
$U_{\bar\sqcup}=\prod_{\ell\in\bar\sqcup}U_\ell$.
\noindent
This model is self-dual: its partition function
\eq
Z(\beta_G,\beta_I)=\sum_{\{\sigma_i=\pm1,~U_\ell=\pm1\}}
e^{-S(\beta_G,\beta_I)}\en
fulfils, in the thermodynamic limit, the functional equation
\eq
Z(\beta_G,\beta_I)=(\sinh 2\beta_G \sinh 2\beta_I)^{\frac 32 N}
Z(\tilde{\beta}_I,\tilde{\beta}_G)
\en
with   $\tilde\beta=-\um\log(\tanh \beta)$.

The phase diagram of this model has been studied long ago
\cite{js}. There is an unconfined region surrounded by lines of phase
transitions toward  the Higgs phase and its dual.
These lines are second order until they are near each other and the
self-dual line, where first order transition occurs.
We want to show that large Wilson loops in the dual Higgs phase obey
the area law.

Using the method of Fortuin and Kasteleyn (FK) \cite{fk} we can map
this system  into a percolation model, rewriting $Z$ as
\eq
Z=e^{-\beta_IN}\sum_{\{U_{ij}\}}
e^{ \sum_{\bar\sqcup}U_{\bar\sqcup}}
\sum_G^\prime v^{n(G)}2^{c(G)}
\en
with $v=e^{2\beta_I}-1$,  $G$ denotes a subgraph of the lattice made of
$n(G)$  links, called active bonds, and $c(G)$ is the number of its connected 
components, called FK  clusters. The only difference between the pure
gauge theory and the one coupled to matter is that in the former
 the $G$ subgraphs are arbitrary, while in the latter the allowed $G$'s
are  subjected to a constraint:  the number of negative links 
($U_{ij}=-1$) along {\sl every} circuit made with active bonds must be 
even, as it is easily checked. Put differently,
no frustration is allowed in $G$, hence no $Z_2$ magnetic flux
can pass through the circuits of $G$. This can be rephrased by saying
that {\sl the FK clusters behave as pieces of super-conducting matter}.
On the other hand,  inserting  a Wilson loop $W(R,T)$ in 
the vacuum corresponds to creating an unit of $Z_2$ flux in the dual version.
Thus, owing to the above super-conducting property, no FK cluster can
be linked with it. Actually, using  the methods of ref.\cite{cg}, one easily
gets the following exact identity 

\eq \bra W(R,T)\ket_{\beta_G,\beta_I} = \bra \varpi(R,T)
\ket_{\tilde\beta_I,\tilde\beta_G}
\label{prj}
\en
where $\varpi(R,T)$ is a projector on the $G$'s
defined as follows
$$ \varpi(R,T)=\cases{0&if some FK cluster is linked to $W$ \cr
1&if no FK cluster is linked to $W$.}$$
Thus the Wilson loop is a sort of cluster counter. It should be mentioned the
surprising analogy of the circuts of the FK clusters with the center 
vortices \cite{ya},\cite{kt}; note however that they have little to do
with the thin vortices associated to the sign of the plaquettes; they
are rather describing new degrees of freedom related to the gauge
variables through a highly non-local duality transformation and live
on links of the dual lattice. They play a relevant role in providing
us with a very efficient disordering mechanism which leads to an area
law for Wilson loops. 

For a large Wilson loop, finite clusters  can link with it  only
along the loop perimeter. Therefore they contribute only to the perimeter
term of eq.(\ref{asymp}). However, owing to the duality transformation
involved in  eq.(\ref{prj}), the Wilson
loop in the {\sl dual} Higgs phase is dominated by an infinite, percolating,
FK cluster characterizing the Higgs phase. It is precisely this
percolating cluster which leads to the
area law for large Wilson loops, as we wanted to show.

\vskip -3. cm
\vskip -4. cm
{\hskip -3.5 cm
\includegraphics[height=50pc,width=30pc]{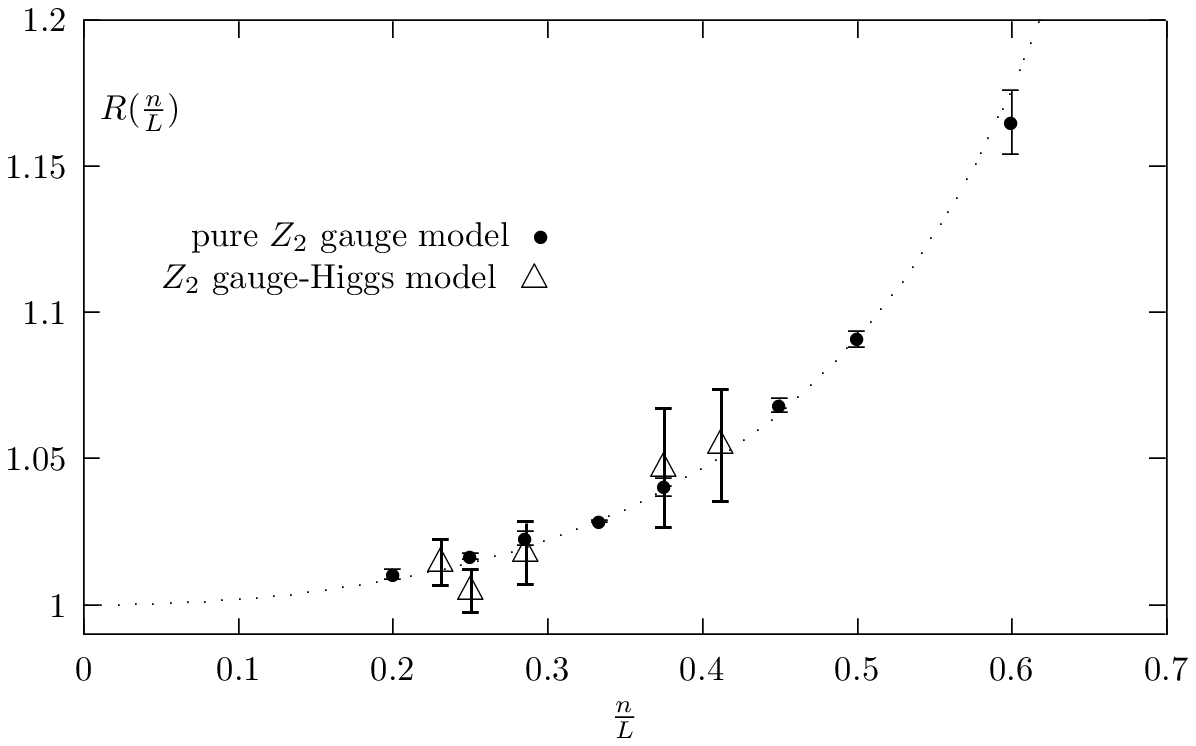}}
\vskip -7.cm
\vskip -1.5 cm
\noindent Figure 1. Universal shape effects in Wilson loops.

\vskip .3 cm
Thus, in the dual of the Higgs phase, even if the ground state  potential
necesssarily flattens out, the area law behaviour of large Wilson
loops is inavoidable.    
As a consequence we can give a precise definition of confinement
phase: it is the one where the area law holds true. In that phase the
string tension $\sigma$ has an unambiguous definiton as the
coefficient of the area term of eq.(\ref{asymp}).

As a check of these results, we have estimated  $\sigma$ in this
gauge-Higgs model in a lattice of size $L^3=40^3$ at $\beta_G=0.75245$
and $\beta_I=0.16683$. We measured all the squared Wilson loops 
$\bra W(R,R)\ket$ with $10< R\le 20$ applying a very powerful 
algorithm already used in the pure gauge theory \cite{gv}, \cite{sew} 
and based on eq.(\ref{prj}). We found
$\sigma=0.00828(26)$.

It also interesting to uncover the universal shape effects produced by
the quantum fluctuations of th string. Actually
the quantity 
\eq R(n,L)=\frac{\bra W(L+n,L-n)\ket}{\bra
  W(L,L)\ket}e^{-\sigma n^2}~~,
\en
 according to the asymptotic form (\ref{asymp}), is only a (known)
 function of the ratio $\frac nL$, does not contain any adjustable
 parameters, nor it depends on the gauge group or on the nature of the
 matter fields. The function $R(\frac nL)$  is plotted in Fig.1 
(dotted line) along with the data of the present study and those of 
pure $Z_2$ gauge theory taken form ref. \cite{sew}. 

\end{document}